# Identified hadron spectra from pp, dA and AA collisions

***Julia Velkovska***

*Department of Physics and Astronomy, Station B 1807,*
*Vanderbilt University, Nashville, TN 37235*

**Abstract.**
The experimental data for identified hadron spectra from *pp, d+A* and *AA* collisions are reviewed. Three regions with different dominant production mechanisms are considered: soft region ($p_T$ < 2 GeV/c) – with large degree of collectivity and thermalization, hard particle production which exhibits jet-quenching in Au+Au collisions at RHIC and intermediate region (2< $p_T$ <5 GeV/c) with distinct baryon dynamics. Cronin effect, nuclear modification factors and jet-like correlations are studied with the goal of understanding the baryon dynamics at RHIC.

## 1. Introduction

The primary goal of experimental relativistic heavy ion physics is to produce and study a de-confined state of nuclear matter: the quark-gluon plasma (QGP). Obtaining single inclusive identified hadron spectra is the first step in deriving a large number of observables. Since heavy ion collisions at relativistic energies involve processes on a variety of energy scales, different parts of the hadron spectra are dominated by different processes. At low $p_T$ (< 2 GeV/c) where the bulk of the particles are produced, collective phenomena and the chemical properties of the system determine the particle yields and ratios. Analysis of the spectra allows the extraction of kinetic and chemical freeze-out temperatures, collective flow velocity and the baryon chemical potential $\mu_B$. Thus the position on the phase diagram can be experimentally determined and then related to the possible QGP formation. The observation of collectivity is also a necessary, although not sufficient condition for the confirmation of QGP.

Above $p_T$ ~2 GeV/c, hard scattering processes become increasingly important. They provide a sensitive tool for probing the produced medium. After the hard-scattering, a colored object (the hard-scattered quark or gluon) traverses the medium produced in the collision and interacts strongly. As a result, it is expected to lose energy via induced gluon radiation. This phenomenon, known as jet-quenching, manifests itself as suppression in the yield of high-$p_T$ hadrons, when compared to the production in *pp* collisions scaled by the nuclear thickness function $T_{AB,}$ which accounts for nuclear geometry. The suppression is measured in terms of the nuclear modification factor $R_{AA} = \sigma^{AA} / T_{AB}\, \sigma^{pp}$. Hadron spectra from *pp* collisions serve as a reference which is free of nuclear effects and will be discussed here in this context. It is also important to distinguish nuclear effects from the initial state (cold nuclear matter) and the final state (supposedly QGP). At RHIC, *d+Au* collision at the same center of mass energy as the *Au+Au* collisions serve this purpose. The Cronin effect, which is usually attributed to initial state multiple scattering is studied in detail.

At this conference, it has been recognized that the medium influences the dynamics of hadronization. The first hint [1] that baryons are formed differently in *Au+Au* collisions at RHIC came with observation that in central collision, near $p_T$ ~ 2 GeV/c proton and anti-proton yields are comparable to the pion yield, contrary to expectations based on the known fragmentation functions. The centrality scaling of the proton and anti-proton yields [2] has also been difficult to fit into the soft+hard description of the hadron spectra. The large number of identified hadron measurements presently available [2-6] makes possible the detailed study of mass effects versus baryon/meson effects. It has become evident that the observed phenomena are specific to nucleus-nucleus collisions and can be linked with the number of quarks in the hadron. Resolving the "baryon anomaly" is the main focus of this talk. The presentation is experiment driven. After a brief survey of soft physics (Section 2) and hard physics (Section 3), I attempt a systematic review of intermediate $p_T$ results in *Au+Au* and in *d+Au* collisions (Section 4). Theoretically, little is known about the flavor dependence of Cronin effect and the specifics of hadronization both in the presence or absence of the QGP medium. The hope is that experimental surveys like this will help guide the theory beyond the description of "pions only". This paper concentrates on the mid-rapidity data from RHIC and the discussion is mainly focused on the identified hadron spectra.

## 2. Soft particle production

In relativistic collisions between nuclei, the bulk of the particle production happens through processes in which the momentum transfer between the interacting nucleons is small (soft collisions). Since QCD has strong coupling at large distances, perturbative methods (pQCD) can not be applied to calculate the cross section. At the microscopic level, only phenomenological descriptions are available. From the other hand, statistical and hydro-dynamics approaches work well [7, 8].

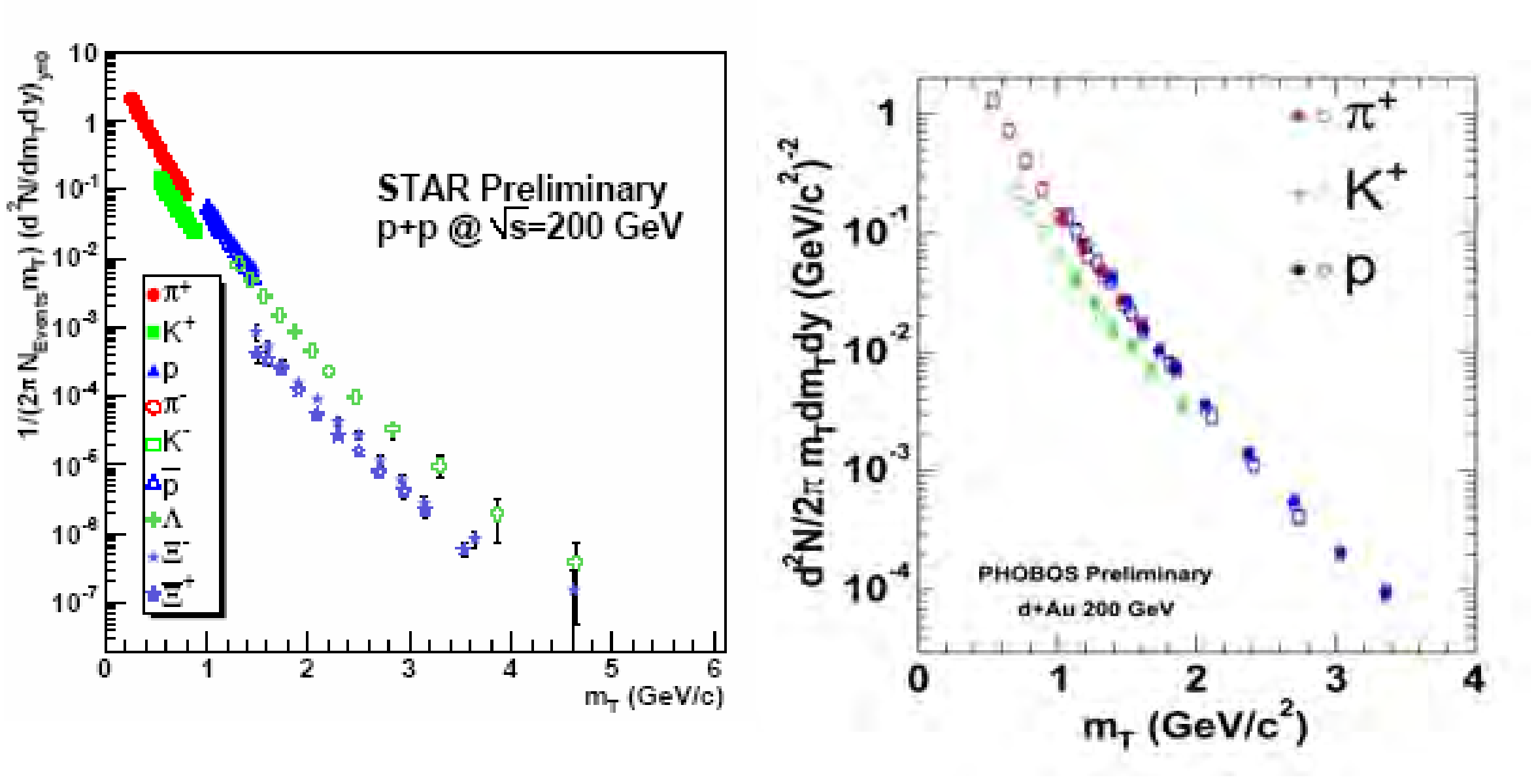

**Figure 1.(left) Identified hadron spectra measure in pp collisions at RHIC by the STAR collaboration [9]; (right) PHOBOS measurement of π, K,p produced in d+Au collisions [10].**

The main feature of the hadron spectra at low $p_T$ is the exponential fall-off typical for thermal production. In this $p_T$ range, the particle mass affects the slope of the spectra. It is customary to plot the invariant yields as a function of the transverse mass $m_T = \sqrt{(p_T^2 + m^2)}$. Then, for *pp* and *d+Au* collisions, the slopes for different particle species become equal. The phenomenon is known as "$m_T$–scaling". Figure 1 shows identified hadron data obtained at $\sqrt{s}$ = 200 GeV. We see that it is possible to rescale the spectra of different particle species by an arbitrary factor, such that all hadrons are described by a universal function, however the spectra do not scale with $m_T$ in an absolute way.

Heavy ion collisions are different. The system develops collective radial flow. The common flow velocity is reflected in the mass dependence of the $m_T$ slopes. The higher the particle mass, the larger is the inverse slope as demonstrated in Fig. 2. Note that the spectra have been (arbitrarily) rescaled at the $m_T = 2$ GeV point in order to facilitate the shape comparison. As $m_T$ increases, the influence of the mass becomes less important and all particle spectra converge to the same slope independent of mass. It is important to realize that at RHIC the inverse $m_T$ –slopes are not constant (Fig. 2 bottom) and thus one can not describe the spectral shapes by just one number, as it has been done for the SPS data [11]. The kinetic freeze-out conditions: freeze-out temperature $T_{fo}$ and transverse flow velocity $\langle\beta_T\rangle$ are best extracted by a full model hydrodynamic calculation. Since these parameters are anti-correlated, the measurement of the spectra alone may not be sufficient to constrain the models. Measurements at very low $m_T$ (filled points in Fig. 2), improve the sensitivity to the flow parameters. A common experimental approach is to use blast-wave parameterization to derive the kinetic parameters [12]. An open question is whether or not all hadrons have the same freeze-out surface and if the situation is the same at SPS and at RHIC. Fits to the spectra consistently give chemical freeze-out temperatures of the order ~ 170 MeV [7], kinetic freeze-out $T_{fo}$ ~ 110 MeV, and $\langle\beta_T\rangle$ ~ ½ c [8]. These temperatures are above the expected critical temperature, both at SPS and RHIC.

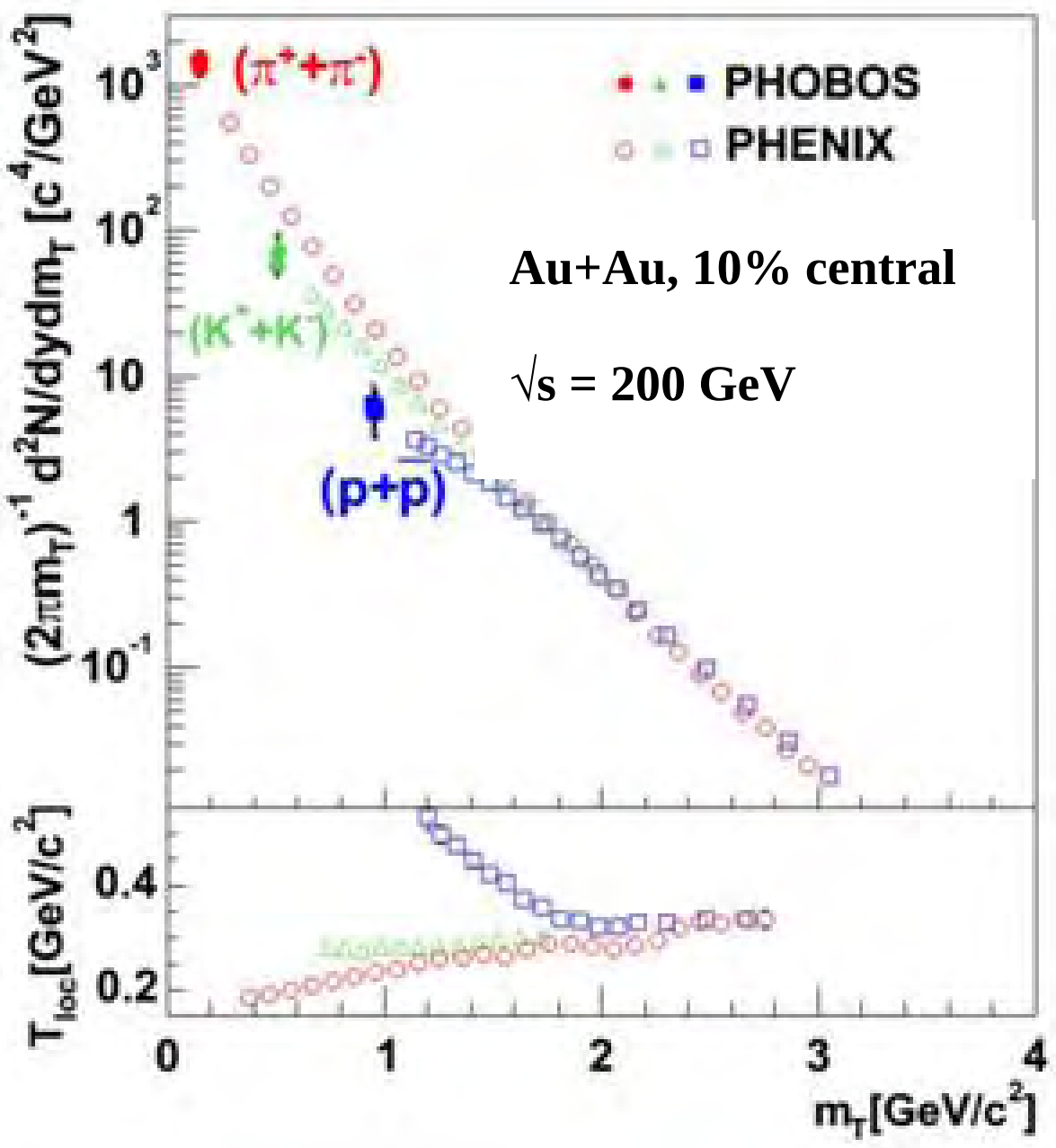

**Figure 2. Open points: $M_T$ spectra for p,K,p for 0-10% central Au+Au collisions (PHENIX [13]). Filled points – PHOBOS [10]. Note: Spectra are rescaled to match at $m_T = 2$ GeV/$c^2$ . Lower panel – local inverse slope.**

## 3. Hard scattering

At high $p_T$, the spectral shapes deviate from thermal production and develop a characteristic power-law tail. The invariant pion yields in *pp* collisions [14] can be described by NLO pQCD calculations over many orders of magnitude (Fig.3).Thus, spectra obtained in elementary collisions serve as a well-calibrated reference for the measurement in nucleus-nucleus collisions. At RHIC, a suppression of hadrons with high transverse momentum was discovered [15]. In addition, the measurement of azymuthal correlations of hadrons revealed vanishing back-to-back correlations in central *Au+Au* collisions [16]. The suppression was observed to increase from peripheral to central collisions and is stronger for pions than for the inclusive charged hadrons. Figure 4 shows the nuclear modification factor $R_{AA}$ for the 10% central collisions. The comparison in the different channels reveals two features: 1) a pronounced difference between identified pions and charged hadrons at $p_T < 5$ GeV/c; 2) constant suppression (~ factor 5) above. The constancy of the suppression with $p_T$ was not originally expected from the theory. For a more detailed discussion, the reader is directed to the hard-probes overview presented at this conference [17].

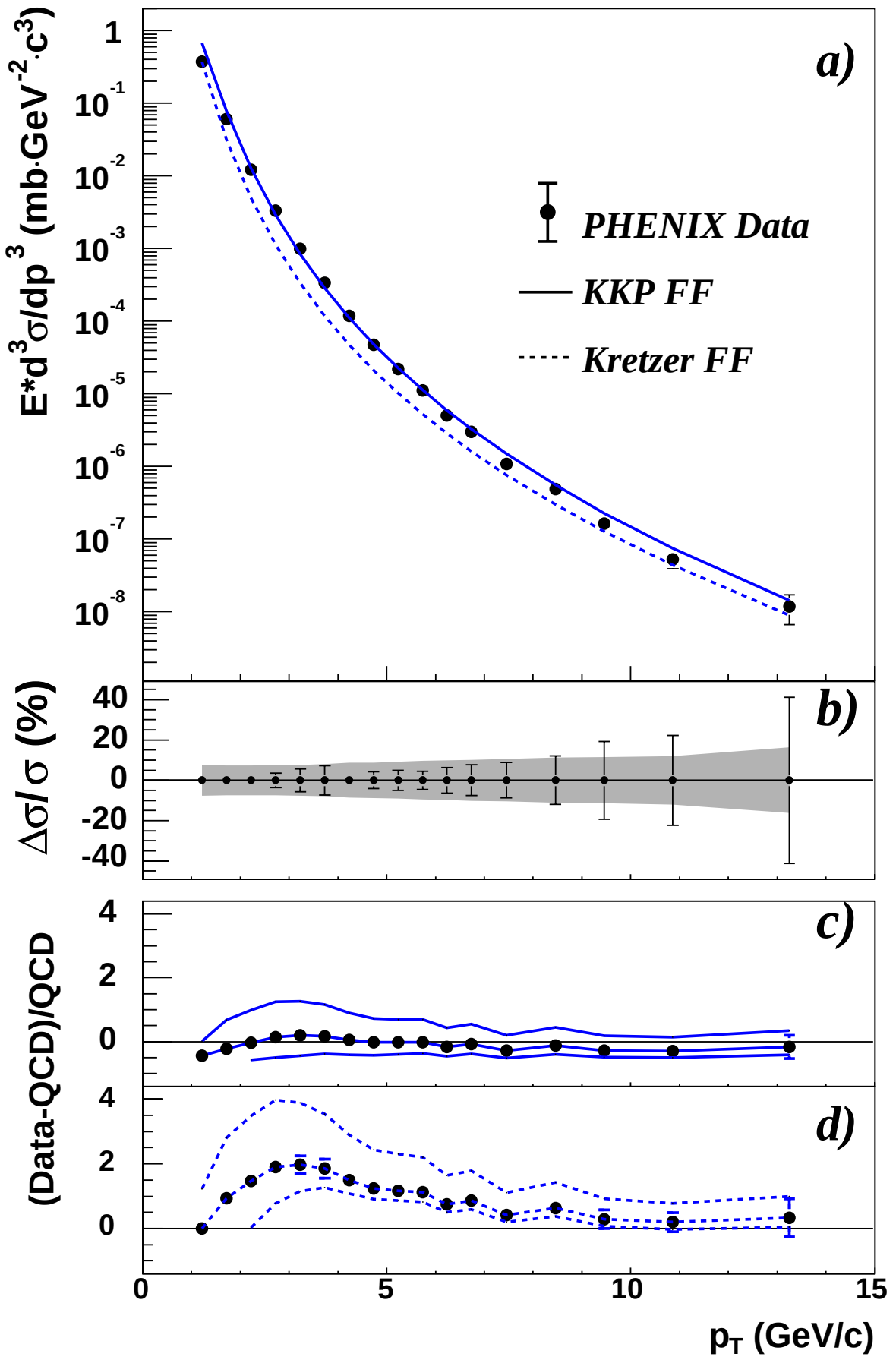

Figure 3. (a) Invariant cross section for $\pi^0$ production in pp collisions at RHIC [14] (b) systematic errors (c) and (d) – data to theory comparisons as labeled in the figure.

This paper is mainly focused at the intermediate $p_T$ identified hadrons. As it will be discussed in Section 4, the different suppression patterns in Fig. 4 are due to the unusually large proton and anti-proton contribution to the charged hadron spectra.

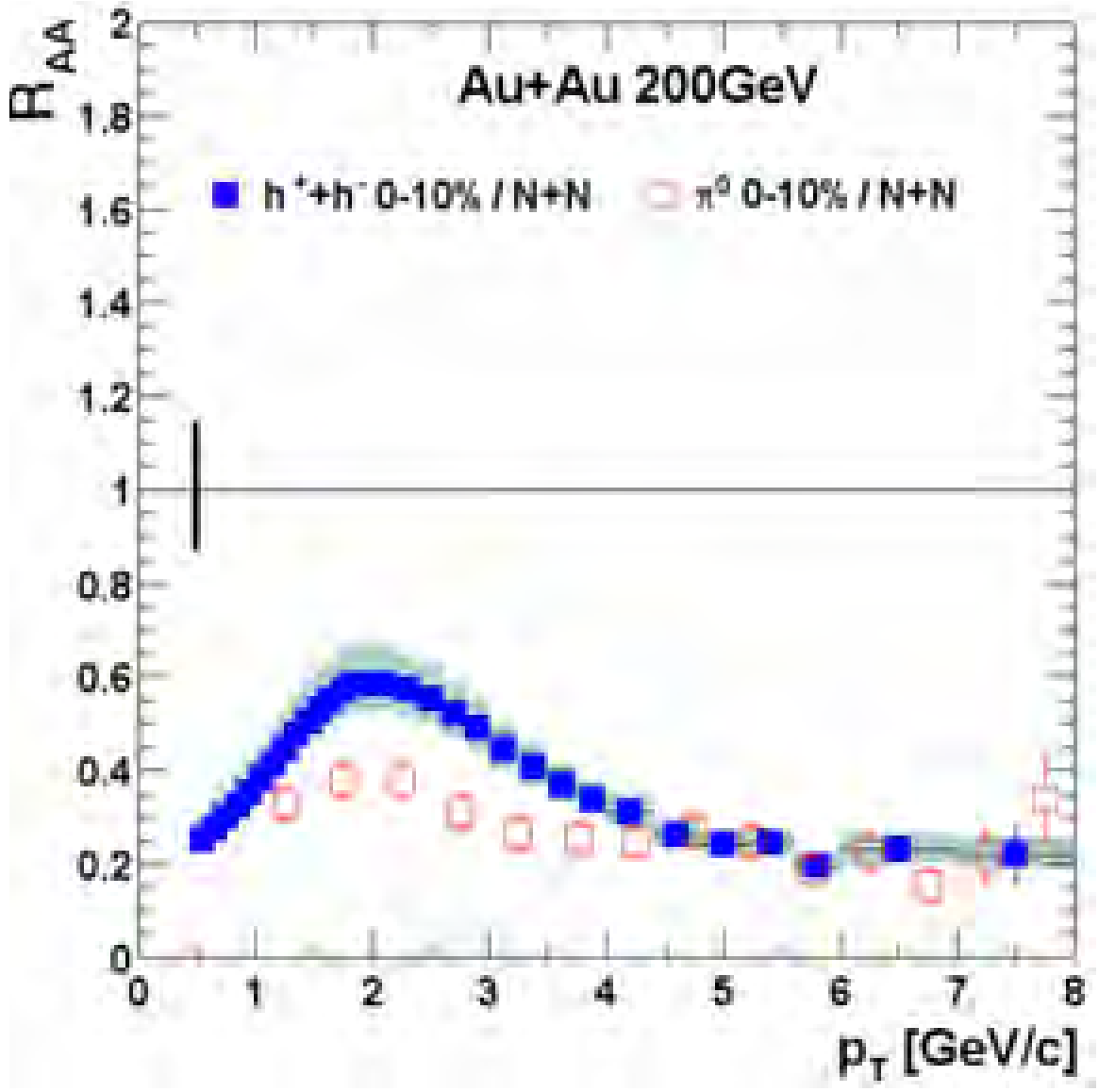

**Figure 4: Nuclear modification factors $R_{AA}$ in 10% central Au+Au collisions measured by PHENIX for $\pi^0$ (open squares) and inclusive charged hadrons (filled squares) .**

The RHIC high-$p_T$ data shows features that are qualitatively different from the observations at SPS, where large Cronin enhancement was observed at intermediate $p_T$. To confirm the jet-quenching scenario, it was necessary to exclude possible modifications caused by cold nuclear matter. The control experiment, *d+Au* collisions, was performed and the suppression of hadrons and the back-to-back azymuthal correlations were shown to be a property of the medium produced in *Au+Au* collisions[18].

## 4. Intermediate $p_T$ : the effect of the medium on hadronization

At RHIC in central *Au+Au* collisions, protons and anti-protons constitute almost half of the charged hadron yield in the region 2< $p_T$ <5 GeV/c. This behavior is centrality dependent. Figure 5 (left two panels) quantifies the effect. The (anti)proton/pion ratio is in peripheral

*Au+Au* collisions is consistent with the ratios measured in *pp* and in $e^+e^-$ collisions. However, as you go from peripheral to central collisions, the ratio grows up to a factor of 3 above the expectations from normal jet fragmentation. Enhanced baryon production is also observed in the strange particle sector. Fig. 5 (right) shows the $\Lambda/K^0$ ratios in *pp* and in centrality selected *Au+Au* collisions. The enhancement of the baryons over mesons that increases with centrality is evident here as well. However, there are two additional features: 1) the ratios in peripheral *Au+Au* collisions are above those in *pp* collisions ; 2) While the (anti)proton/pion ratios level off and do not come down in the measured range, the $\Lambda/K^0$ ratios have a peak that moves towards lower $p_T$ as you go from central to peripheral collisions. The former maybe related to the strangeness content and is not well understood at this time. The latter was predicted by theories that invoke baryon junctions to explain the data [19]. It is expected in other models as well, as it is a necessary feature, if one wants to recover the vacuum fragmentation functions at sufficiently large $p_T$.

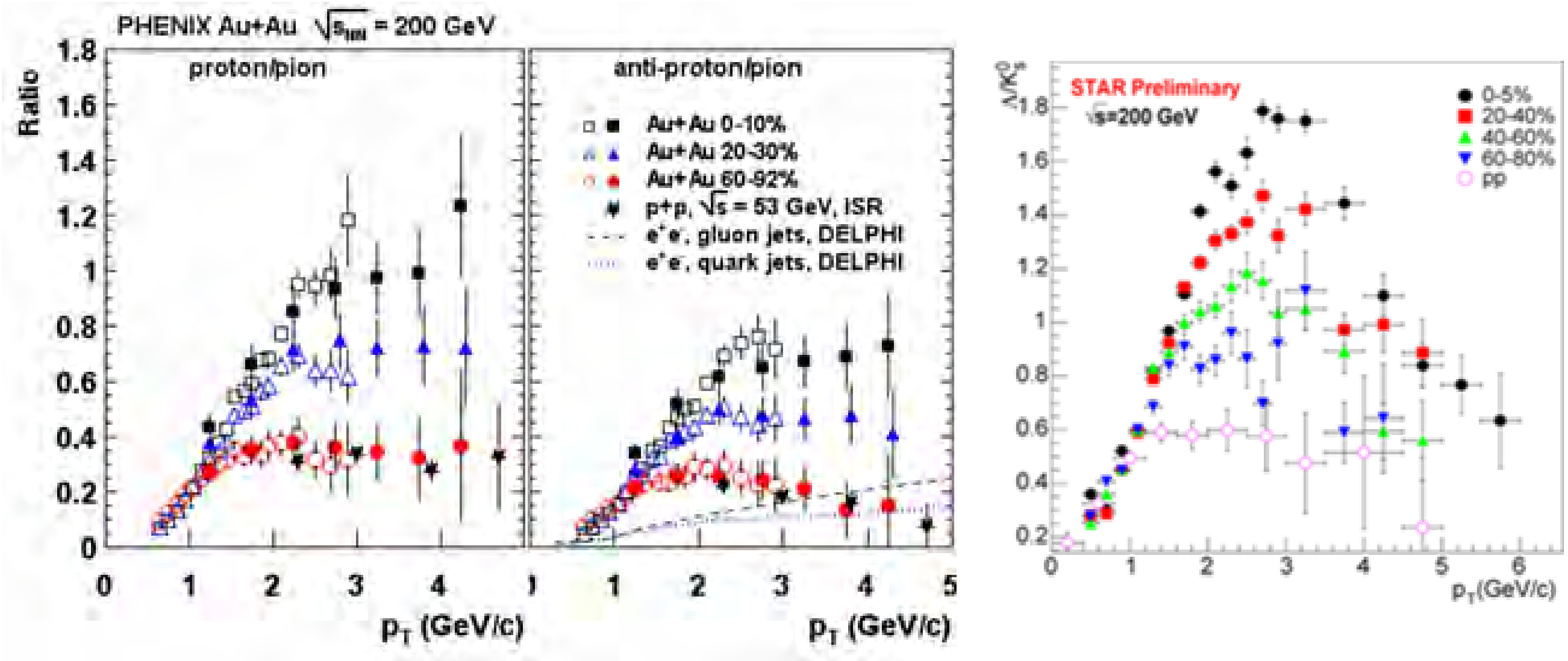

**Figure 5: Evolution of baryon/meson ratios with centrality in *Au+Au* collisions at RHIC. Left two panels are from publication [2] of PHENIX. Right: preliminary data from the STAR experiment [4].**

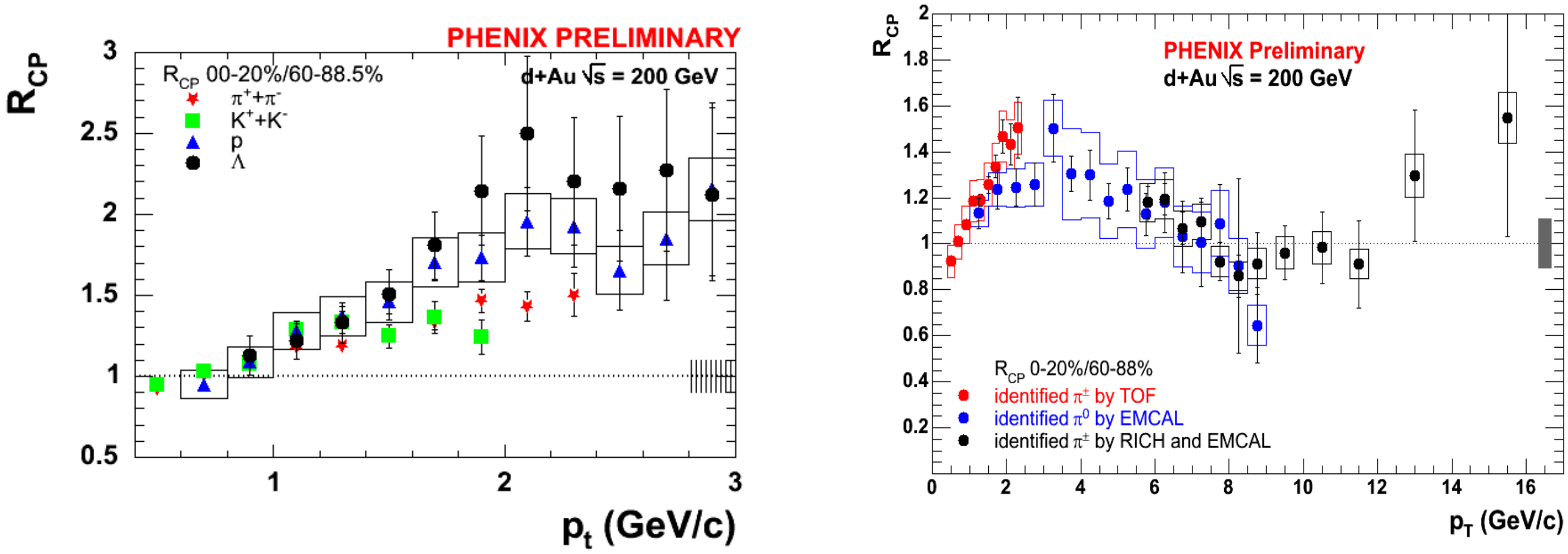

**Figure 6: $R_{cp}$ (0-20%/60-88%) measured for π, K, p and Λ in *d+Au* collisions at RHIC. (Right) Compilation of $R_{cp}$ for charged and neutral pions, combining several different detector technologies in PHENIX.**

Enhancement of proton production compared to pions at intermediate $p_T$ has been previously observed when comparing *pA* collisions in different nuclear targets [20]. The Cronin enhancement in *p+W* collisions is factor of ~ 2 larger for protons than for pions for 3< $p_T$ < 6 GeV/c. It is important to establish if the observed effects in *Au+Au* collisions at RHIC can be understood in terms of initial state $p_T$ – broadening. Theoretically, Cronin effect is still not well understood [21]. Most theories try to describe the effect for pions, if they carry through to the hadronic stage at all. The particle species dependence is usually not addressed. Here, I summarize the currently available experimental information from RHIC. In Figure 6, the effect is quantified in terms of the central to peripheral ratios $R_{cp}$, each scaled by their respective number of $N_{coll}$ to account for nuclear geometry. These ratios carry similar information to the nuclear modification factors $R_{dA}$ (measured for π,K,p in [22]).. Figure 6 (left) shows the particle species dependence of the Cronin effect at RHIC energies: protons and lambdas show larger enhancement than pions and kaons, but the effect is much smaller than at lower energies. Strangeness content does not seem to affect the behavior. The enhancement grows from peripheral to central collisions (see e.g. [5,6]), but tends to saturate after several collisions [6]. Figure 6 (right) is for pions only and aims to determine the $p_T$ range of the enhancement. It shows that the enhancement is limited to 6-8 GeV/c in $p_T$.

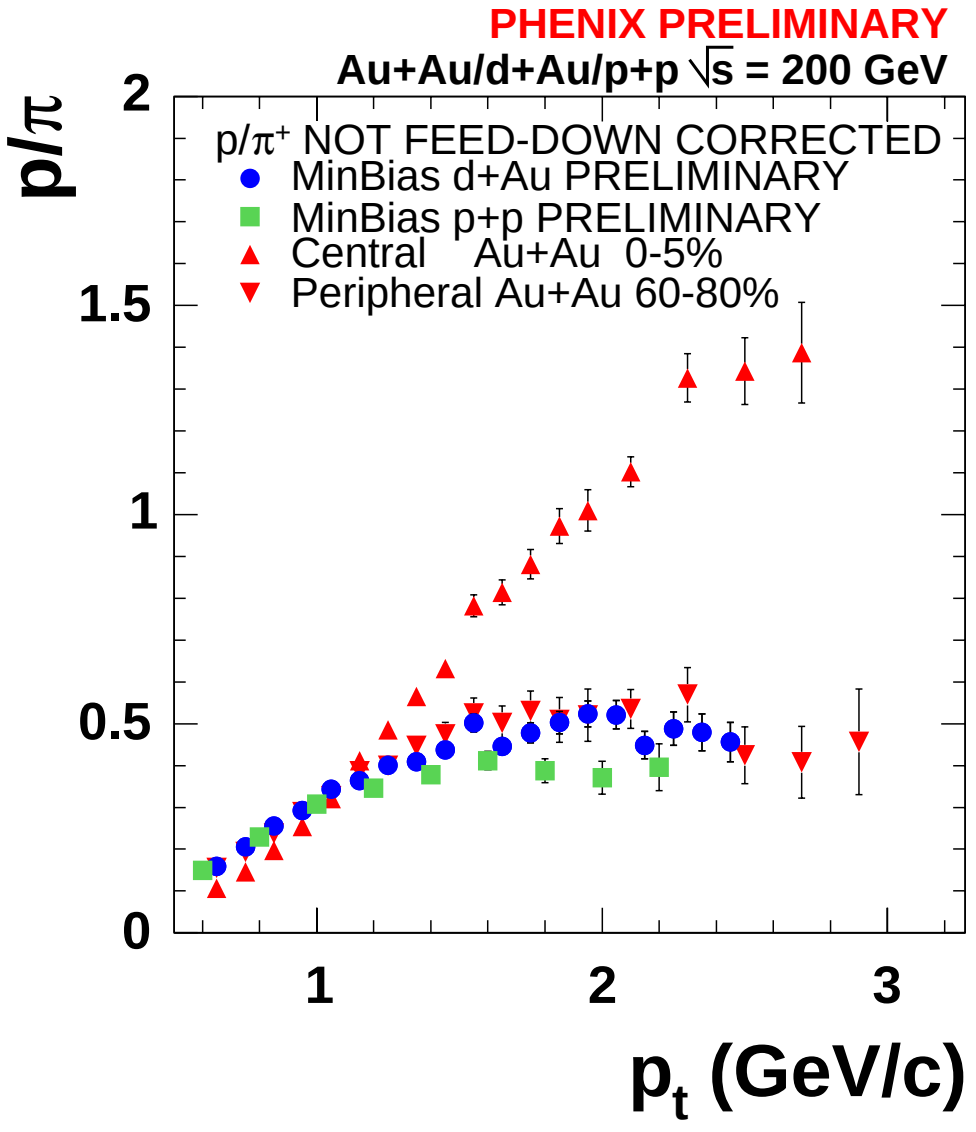

**Figure 7. p/π ratio for AuAu, dAu and pp collisions at RHIC [6].**

In Figure 7 we come back to the p/π ratio. This time we compare three colliding systems: *Au+Au* central and peripheral (shown with up and down triangles, *d+Au* – circles and *pp* – (squares). The different Cronin enhancement reflects in about 20% difference between the *d+Au* and *pp*. The central *Au+Au* result stands out as dramatically different from the others. Clearly, the observed baryon enhancement is a property of the medium produced in central *Au+Au* collisions. The effect present in *d+Au* collisions is of much smaller magnitude.

Another surprising feature of the baryons produced at intermediate $p_T$ is that they do not participate in jet quenching. In Figure 8, protons (left), Λ,Ξ and Ω (right) are not suppressed. Again, the proton measurement levels off above $p_{T\sim}$ 2 GeV/c and does not come down within the measured range. The measurement of strange baryons has larger $p_T$ range and shows that above $p_T$ ~ 5 GeV/c, the particle species dependence disappears. The same (indirect) information for the protons is contained in Fig. 4. One proposed explanation for the observed effect has been the hydrodynamics boost that extends the $p_T$ range of soft production for heavier particles [8]. To test this hypothesis, we compare the scaling behavior of ϕ mesons to that of protons (Fig. 8 left). If mass is the mechanism that makes pions and protons different, then we expect that ϕ and p should behave in the same way, since they have very similar masses (moreover $m_\phi > m_p$ ). The experiment shows the opposite: ϕ production is suppressed, unlike the production of protons and anti-protons. Similarly, the heavy $K^*$ meson (Fig. 8 right) shows suppression, while Λ,Ξ,Ω do not ($p_T$<5 GeV/c). These measurements rule out the mass as the source of the proton and Λ enhancement. Clearly, we have an observable that is sensitive to the number of quarks in the hadron!

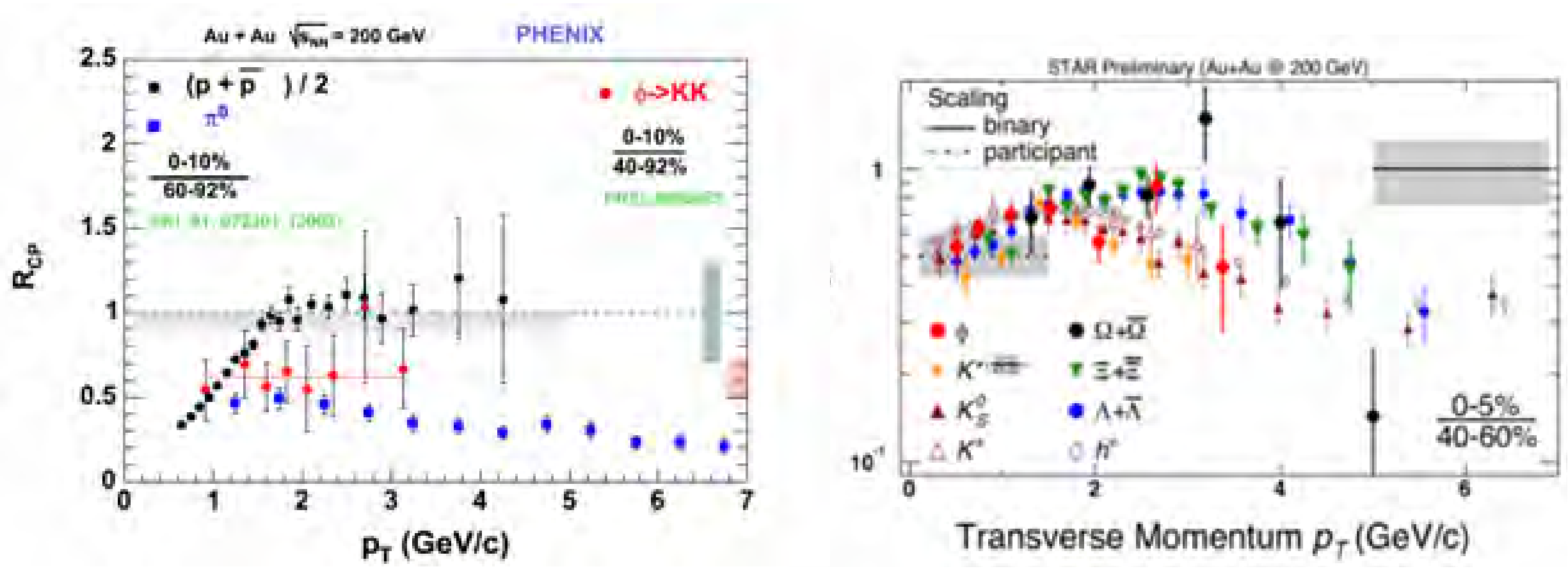

**Figure 8. Binary scaled central to peripheral ratios $R_{cp}$ for identified particles [2-5]. The nuclear modification follows baryon/meson lines independent of mass.**

Quark recombination models [23] have been successful in reproducing the particle yields, scaling behavior and elliptic flow patterns of baryons and mesons. The hot and dense medium produced in *Au+Au* collision influences the process of hadronization. The elliptic flow data is well described in term of partonic flow – maybe the most direct evidence that partonic degrees of freedom are released at RHIC. In this picture, if the baryons come from recombination of thermalized quarks, we do not expect to see jet-like correlations of leading baryons with other hadrons. This is not the case, as demonstrated in Fig. 9 [24].

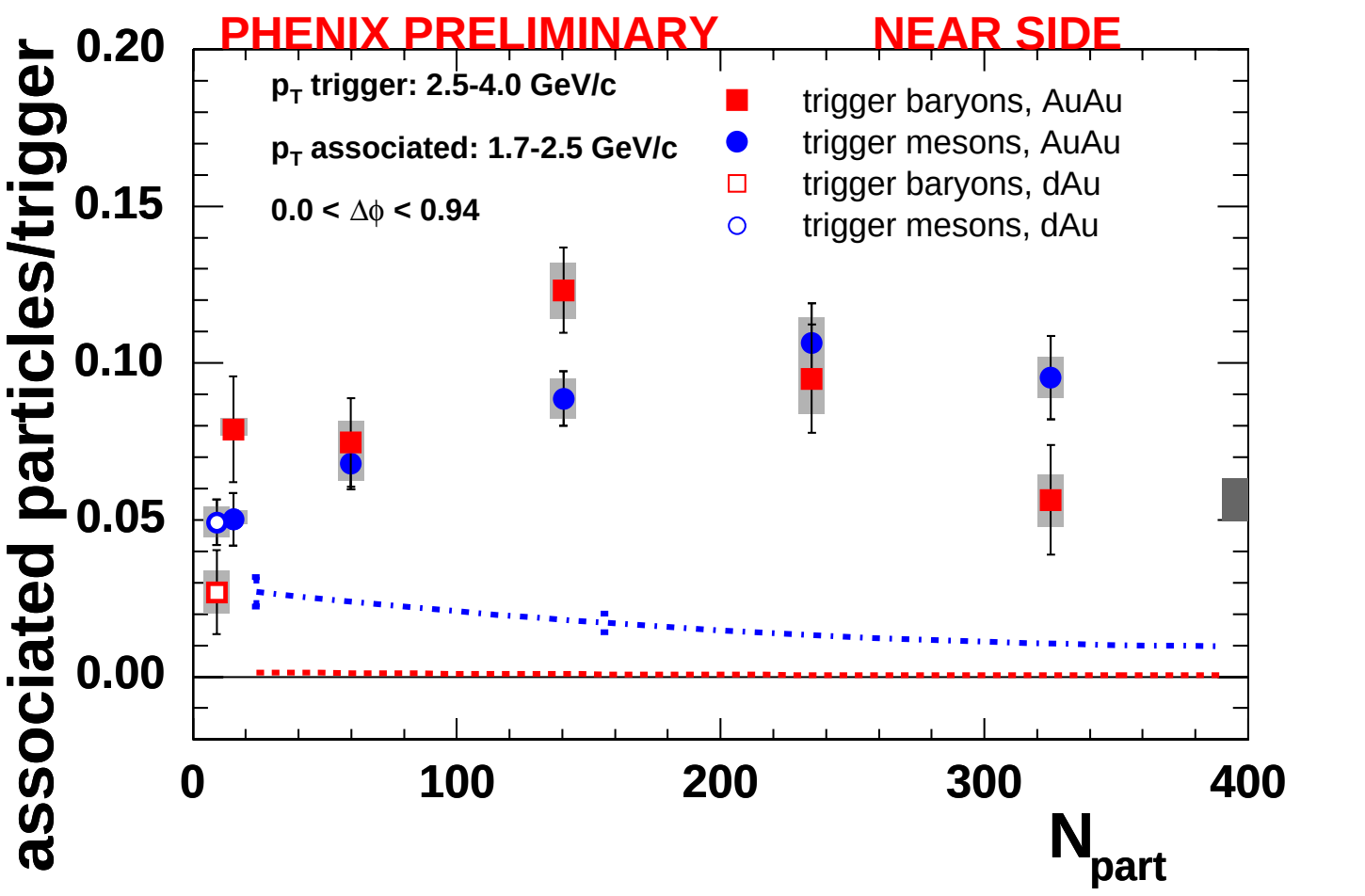

**Figure 9. Associated near-angle yield per identified hadron trigger at intermediate $p_T$. The lines are expectation from thermal recombination theory [23].**

The yields in the near-side peak associated with identified mesons and baryons are compared and studies as a function of centrality and colliding system. Surprisingly, no significant difference is found, although at intermediate $p_T$, baryons clearly do not obey normal jet fragmentation. A possible solution is that recombination works between thermal and shower partons [24]. This, however, would destroy the nice agreement of recombination models with the elliptic flow data. Although, recombination seems a simple and efficient solution to the baryon production at RHIC, it still has experimental challenges to meet.

## 4. Conclusion

Identified hadron spectra carry a wealth of information about the dynamics of heavy ion collisions at relativistic energies. The soft part of the spectra looks thermal and points to significant collective radial flow. Hard scattering processes in *Au+Au* collisions at RHIC reveal unique features. Jet quenching was established with a series of measurements in different colliding systems: *pp*, *d+Au* and *Au+Au*. At intermediate $p_{T,}$ baryons were found to dominate the hadron spectra, contrary to know fragmentation functions. Their spectra in *Au+Au* collisions are not quenched, although they exhibit significant jet-correlations. Cronin effect is stronger for baryons than for mesons, but can not account fully for the large baryon enhancement in central *Au+Au* collisions. Soft mass effects are excluded by the observation that heavy mesons ($\phi$ , $K^*$) are suppressed like the lighter mesons. The recombination of thermal plus shower quarks seems to be the most probable explanation to the observed baryon enhancement.